\documentclass[footinbib,twocolumn,showpacs,amsmath,amstex,amssymb,mathfonts,superscriptaddress,prl]{revtex4}
\usepackage{graphicx}
\usepackage{color}
\usepackage{bm}

\usepackage{graphicx}
\usepackage{color}
\usepackage{bm}
\usepackage{amsmath}
\usepackage{amssymb}
\usepackage{amsthm}
\usepackage{amsfonts}
\usepackage{multirow}
\usepackage{makecell}
\usepackage{float}
\usepackage{comment}
\usepackage{enumitem}
\usepackage{verbatim}
\usepackage{mathtools}

\usepackage{bbm}

\begin{document} 

\title{Emergent dynamics of scale-free interactions in fractional quantum Hall fluids}
\author{Bo Yang} 
\affiliation{Division of Physics and Applied Physics, Nanyang Technological University, Singapore 637371.}

\date{\today}
\begin{abstract}
We show that even with arbitrarily large cyclotron gap, Landau level (LL) mixing can be dominant with scale-free interaction in a fractional quantum Hall system, as long as the filling factor exceeds certain critical values. Such scale-free interaction with kinetic energy can serve as exact model Hamiltonians for certain composite Fermion or parton states (unlike the well-known TK Hamiltonians where the number of LLs needs to be fixed by hand), and they are natural physical Hamiltonians for 2D systems embedded in higher-dimensional space-time. Even with LL mixing the null spaces of such Hamiltonians (spanned by the ground state and the quasiholes) can be analytically obtained, and we show these are the generalisation of the conformal Hilbert spaces (CHS) to more than one LLs. The effective interaction between the anyons for these topological phases emerges from the kinetic energy of the ``elementary particles", leading to an interesting duality between strongly and weakly interacting systems that can be easily understood via the tuning of parameters in the scale-free interaction. We also propose a novel experimental platform for approximately realising such model Hamiltonians with trion like particles, that can potentially lead to very robust (non-Abelian) FQH phases from two-body Coulomb-based interaction.

\end{abstract}

\maketitle 

The long range bare Coulomb repulsion between electrons comes from one of the fundamental interactions between elementary particles. The inverse quadratic interaction (e.g. $1/r^2$, where $r$ is the separation between the two particles) is a universal force that is scale-free; the quadratic power is fundamentally determined by the three-dimensional universe we live in, from the hydrodynamic description of the gauge Bosons mediating the interaction between charged particles. The long range nature of the interaction tends to complicate theoretical treatments, where singularities need to be properly renormalised. In Fermi liquid such long range interaction typically can be screened by mobile charge carriers\cite{mccomb}, leading to a Yukawa like interaction and Friedel oscillation of electron density\cite{harrison}. 

For strongly interacting condensed matter systems with flat bands (quenched kinetic energy), the quantum fluids of electrons are quintessentially non-Fermi liquid. In two-dimensional electron gas systems with a strong perpendicular magnetic field, flat bands called Landau levels are formed. The fractional quantum Hall effect (FQHE) emerge entirely from Coulomb interaction (with no kinetic energy), when the top most Landau level is partially filled\cite{prange}. Nevertheless, extensive numerical analysis shows single-particle like behaviours emerge from strong interaction, not for electrons, but for emergent phenomenological composite Fermions\cite{jain}. At least for finite size systems, the composite Fermion wavefunctions have large overlap with the Coulomb interaction ground states, thus capturing not only the topological properties of those quantum fluids, but also the non-universal quantitative behaviours quite accurately.

It is important to note that for many FQH phases, there exist model Hamiltonians and model wavefunctions (as exact eigenstates) that give ideal descriptions of such topological phases\cite{prange,simon}. These include the exact quasihole counting and degeneracy, as well as the ground state topological entanglement entropy and spectrum\cite{kitaev,wen,haldane}. Such model Hamiltonians in the form of pseudopotentials are exact toy models for the FQH phase. For the CF wavefunctions or the exact eigenstates of the Coulomb interactions, however, these ideal properties are conjectured to be qualitatively captured but no longer quantitatively exact. In particular, the CF wavefunctions after the lowest LL (LLL) projection generally do not even have parent Hamiltonians\cite{sreejith, yang1}. It is thus interesting that the CF wavefunctions tend to quantitatively agree with the Coulomb eigenstates better than the special model wavefunctions. A more detailed exposition on the microscopic derivation of the CFs without the LLL projection has been discussed in\cite{yang1,yang2}. The questions of why the Coulomb interaction seems to be special for the composite Fermions, and its hidden connection to the model Hamiltonians and the LLL projection, are still poorly understood.

In this Letter, we study the family of scale-free interactions between elementary Fermions (e.g. electrons) and Bosons with the electrostatic interaction $V_\alpha(r)=1/r^\alpha$, confined to a two-dimensional manifold. For integer $\alpha$ this is the Coulomb interaction within a 2D manifold in a universe with $\alpha+2$ spatial dimensions; though in principle $\alpha$ serves as a single continuous tuning parameter for a family of Hamiltonian of the following form:
\begin{eqnarray}\label{ham}
\hat H_\alpha=\sum_i\frac{1}{2m}g^{ab}\hat\pi_a\hat\pi_b+\int d^2r_1d^2r_2V_\alpha\left(|\vec r_1-\vec r_2|\right)\hat\rho_{\hat r_1}\hat\rho_{\hat r_2}
\end{eqnarray}
Here $g^{ab}$ is the metric describing the effective mass tensor with $m$ being the effective mass, $\hat\pi_a=\hat P_a-e\hat A_a$ is the canonical momentum from the minimal coupling of the external vector potential $\hat A_a$ with the uniform magnetic field $B=\epsilon^{ab}\partial_{\hat r^a}\hat A_b$ and the electron charge $e$; $\hat\rho_{\hat r}=\sum_i\delta^2\left(\hat{r_i}-\vec r\right)$ is the bare electron density operator. 

We show even in the experimentally relevant limit of $\hbar\omega_c\to\infty$ where $\omega_c=eB/m$ is the cyclotron frequency, the low energy physics could naturally involve multiple LLs and the Trugman-Kivelson (TK) type interaction Hamiltonians \cite{tk} are related to the special cases of $\hat H_\alpha$ with $\alpha=2k, k\ge 1$. Thus in the large $\omega_c$ limit, strong LL mixing can still occur when the filling factor \emph{exceeds critical values}, and the low energy dynamics (e.g. ground state and quasiholes) can be exactly obtained even with LL mixing. Indeed from our understanding of the TK Hamiltonian\cite{rezayi,seidel}, Eq.(\ref{ham}) is the \emph{exact physical model Hamiltonian} with proper values of $\alpha$ for a number of the unprojected CF and parton states with no additional assumptions. This is not the case for the TK Hamiltonian where the number of LLs needs to be \emph{manually determined}\cite{footnote1}. The conjectured adiabatic connection to the LLL Coulomb interaction physics for these CF and parton states comes from tuning a single parameter $\alpha$ while keeping the cyclotron energy large. Based on those theoretical results we propose novel experimental platforms where model Hamiltonians and non-Abelian physics from two-body interactions can be potentially realised.

It is useful to first confine the Hilbert space to the LLL, and look at the effective interaction from $V_\alpha(r)$ by expanding its Fourier transform in the basis of Haldane pseudopotentials. The effective interaction in the momentum space is:
\begin{eqnarray}\label{fourier}
 \tilde V_\alpha(q)\sim q^{\alpha-2}e^{-\frac{1}{2}q^2}=\sum_{i=0}^\infty c_{\alpha,i}V^{\text{2bdy}}_i
 \end{eqnarray}
where the two-body pseudopotentials (PP) are $V^{\text{2bdy}}_i=L_i(q^2)e^{-\frac{1}{2}q^2}$ with $L_i(x)$ being the $i^{\text{th}}$ Laguerre polynomial and $c_{\alpha,i}$ being the PP weights. The Coulomb interaction is the well-known case of $\alpha=1$, but Eq.(\ref{fourier}) is only valid for $\alpha<2$; for $\alpha\ge 2$ the Fourier transform diverges. A more careful calculation shows that as $\alpha$ approaches $2$ from below, only $c_{\alpha\to 2, 0}$ is divergent, while all other PP weights are finite. More generally as $\alpha\to 2k$, $c_{\alpha\to 2k,i\le k-1}\to\infty$ (thus equivalent to a divergent $\nabla^{2k-2}\delta^2\left(\vec r_1-\vec r_2\right)$) while $c_{\alpha\to 2k,i>k-1}$ are all finite. As $\alpha$ increases, the interaction diverges more strongly at short distances, and it is the leading PPs that become divergent first, forbidding two particles from having small relative angular momenta.
  
\textit{Density dependent LL mixing--}
The simplest example is the Bosonic system with $\alpha=2$, so within the LLL, $c_{2,0}$ is infinity, leading to a hardcore interaction equivalent to the artificial model Hamiltonian (with infinite strength) for the Bosonic Laughlin state at filling factor $\nu=1/2$. With large cyclotron energy, for $\nu\le 1/2$ (or more specifically for the number of magnetic fluxes $2Q\ge 2N_b-2$ on the spherical geometry with $N_b$ the number of Bosons) the ground state(s) are completely within the LLL, which we can set as zero energy states. Thus the null space of Eq.(\ref{ham}) is spanned by the \emph{model} ground state and the quasiholes of the Laughlin $\nu=1/2$ phase\cite{sm}, with the highest density states given by $2Q=2N_b-2$. At low temperature, no LL mixing is induced by the interaction in the limit of large cyclotron energy (see Fig.(\ref{figb}a)). 

Adding another Boson to the highest density quantum fluid, however, is not possible within the LLL, since the interaction will send the energy to infinity\cite{footnote4}. At $2Q<2N_b-2$, strong LL mixing is present \emph{no matter how large the cyclotron gap is}. Indeed for $\alpha=2$, the second term of Eq.(\ref{ham}) is equivalent to an \emph{infinite} bosonic TK interaction. Only the second LL will be occupied for $2Q\ge3N_b/2-3$ (see Fig.(\ref{figb}b)). These are known as the exact unprojected Bosonic Jain state at $\nu=2/3$ and its quasiholes\cite{rezayi}. Further increasing the density with $2Q>3N_b/2-3$ leads to the consecutive mixing of higher LLs (see Fig.(\ref{figb}c))\cite{footnote}, giving a series of conjectured non-Abelian FQH states that are well studied in the context of the TK Hamiltonians\cite{mccann,seidel}, but without artificially picking a few lowest LLs by hand. A schematic illustration of different topological phases at larger $\alpha$ is shown in Fig.(\ref{figb}d), which we will also explain in more details later.

\textit{Model Hamiltonians for the composite Fermion and parton theory--}
The family of TK Hamiltonians involving multiple LLs give exact unprojected Jain states and parton states in many cases but it is not obvious how they are relevant to experiments when one cannot just select just a few LLs arbitrarily. The family of Hamiltonians with scale-free interactions given in Eq.(\ref{ham}), however, serves as natural exact model Hamiltonians for many Jain states and parton states. It also offers transparent insights on why such theories are successful for the Coulomb interaction (with $\alpha=1$)\cite{jain}, and why LLL projection in such theories could be conceptually redundant\cite{yang1} and may only be useful for certain numerical computations.

\begin{figure}
\begin{center}
\includegraphics[width=\linewidth]{fig_b.pdf}
\caption{An illustrative example for $\alpha=2$, in a) to c) the circles are orbitals in three different LLs, the solid ones are occupied by Bosons. a). All Bosons can go into the LLL when $\nu\le 1/2$; b). An additional Boson added to the half-filled LLL has to go to the second LL to avoid infinite energy; c). An additional Boson added to filling factor $\nu=2/3$ has to go the third LL; d). A schematic illustration of different CHS in the parameter space of $\alpha$ and $\nu$. Here $L_3\in L_2\in L_1$ are CHS within the LLL, and $S_2\in S_1$ are CHS within two lowest LLs, while $T_1$ is the CHS within the lowest 3 LLs\cite{fn}.}
\label{figb}
\end{center}
\end{figure}

This is because with Eq.(\ref{ham}) as the exact model Hamiltonian (e.g. for the Jain state at $\nu=2/3$ or the $221$ parton state at $\nu=1$), it is natural to include more than one LLs. As we adiabatically tune $\alpha$ to 1, all states are orthogonal though they are mostly within the LLL in the presence of the large cyclotron gap: it is not necessary to project the CF wavefunctions, leading to the spurious non-orthogonality. Certain unprojected CF states will vanish entirely under LLL projection. This corresponds to eigenstates of Eq.(\ref{ham}) that lives entirely within higher LLs, and thus their energies are sent to infinity (as punished by the cyclotron energy).

\begin{figure}
\begin{center}
\includegraphics[width=\linewidth]{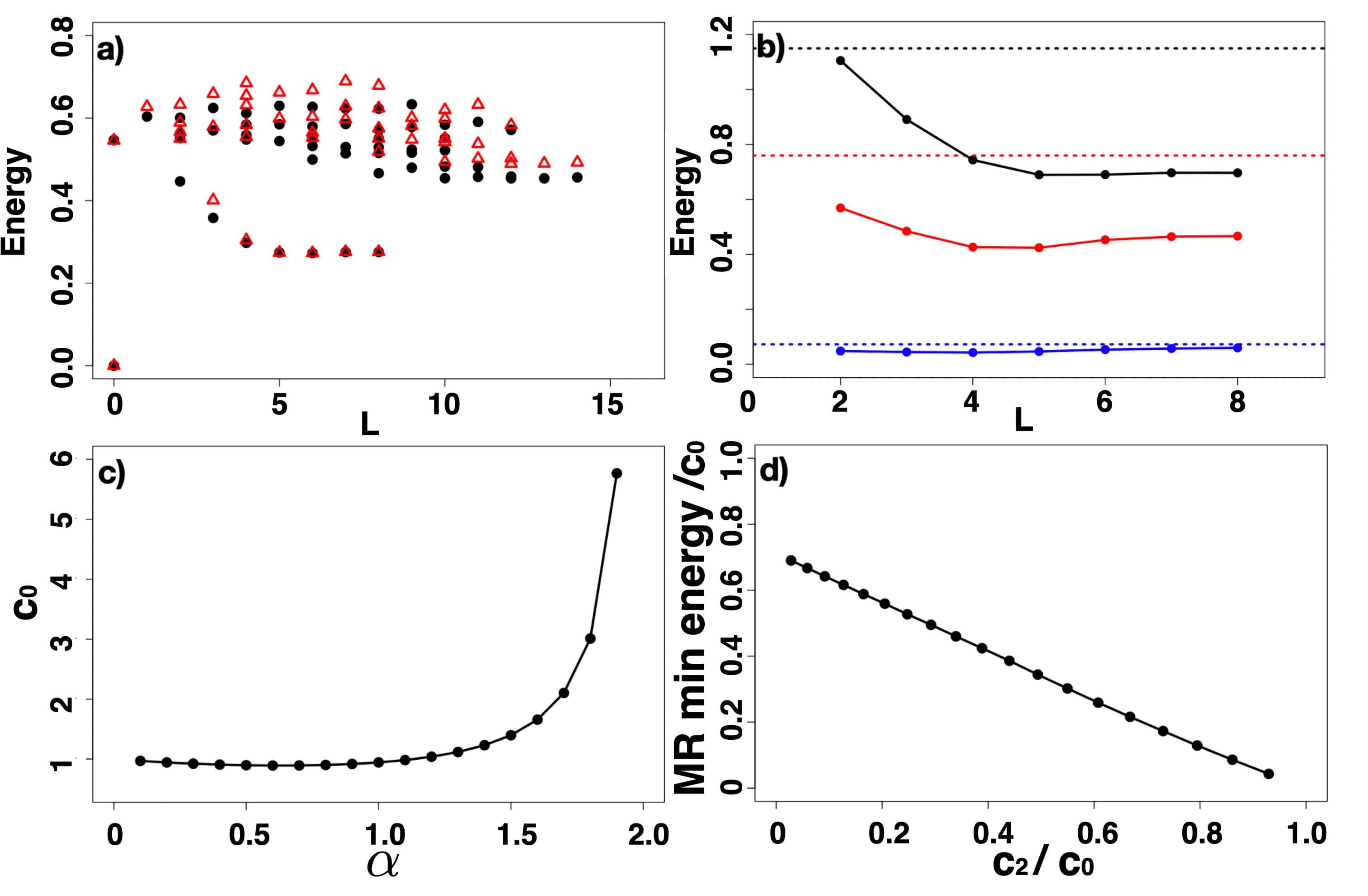}
\caption{a). Energy spectrum on the sphere with $2Q=14$ and $N_b=8$, with the black plot from exact diagonalisation at $\alpha=2-\epsilon$, and the red plot at $\alpha=2$, both from Eq.(\ref{ham}). $L$ s the total angular momentum quantum number. b). The normalised magnetoroton mode for different values of $\alpha$, with $\alpha=1.9$(black), $\alpha=1$(red) and $\alpha=0.1$ (blue). The horizontal dotted lines are the bottom of the multi-roton continuum. c). The dependence of $c_{\alpha,0}$ on $\alpha$. d). The dependence of the normalised magnetoron minimum energy on $c_{\alpha,2}/c_{\alpha,0}$ (R-squared error of the linear fit is 0.9997). }
\label{figa}
\end{center}
\end{figure}

Indeed in Fig.(\ref{figa}) we can see in the large cyclotron gap limit, there is numerical evidence that scale-free interactions are qualitatively similar\cite{sm}. This is true for the magnetoroton modes (MR) of the Laughlin state\cite{gmp,yang2} at $\nu=1/2$, both at $\alpha=2$ where the second LL is involved (see Fig.(\ref{figa}a)), and for $0\le\alpha<2$ when only the LLL is involved (see Fig.(\ref{figa}b)). Here the gap of the MR is controlled by both $c_{\alpha,0}$ and the perturbation to $V^{\text{2bdy}}_0$ (i.e. $c_{\alpha,i>0}$). In Fig.(\ref{figa}b) the MR energies are normalised by $c_{\alpha,0}$, so the gap is entirely from $c_{\alpha,i>0}$. It is clear that $c_{\alpha,0}$ approximately approaches unity as $\alpha\to 0$ (see Fig.(\ref{figa}c)), as do all other $c_{\alpha,i}$; there is almost a linear relationship between the normalised gap and $c_{\alpha,2}$, as shown in Fig.(\ref{figa}d) (see caption). Thus as $\alpha\to 0$, the energy scales (e.g. the gap and the bandwidth of MR, etc) approach zero, but the dynamics of the MR does not qualitatively change\cite{footnote5}. This is highly non-trivial, since for small $\alpha$, the perturbation to $V^{\text{2bdy}}_0$ is very large (e.g. $c_{\alpha,i>0}/c_{\alpha,0}\to 1$). Note that at the limiting case of $\alpha=0$, we have a constant interaction, thus all states within the LLL have the same energy. The integer quantum Hall effect (IQHE) thus can also be seen as a special case of the scale-free interaction (i.e. Eq.(\ref{ham}) with $\alpha=0$). 

\textit{Conformal Hilbert Spaces within multiple LL--}The concept and hierarchy of conformal Hilbert spaces (CHS) within a single LL have been discussed extensively with various applications\cite{yang3,yang4,yang5,yang6,yang7}. These are Hilbert spaces in a 2D manifold with emergent conformal symmetry, exact bulk-edge correspondence, containing a particular type of anyons/non-Abelions as ``elementary particles". Thus each CHS is the low energy manifold of a particular topological phase; for scale-free interaction the corresponding CHS is spanned by all the finite energy states. At $\alpha=2$ (or $\alpha=4$), for Bosons (or Fermions) the null space for the scale-free interaction is isomorphic to the CHS of 2 LLs for $\nu\le 2/3$ (or $\nu\le 2/5$), completely agreeing with the CF description for the IQHE of composite Fermions (from each electron attached to two fluxes). For greater filling factor, the null spaces of the scale-free interaction involves more than 2 LLs, and become non-Abelian. 

In Fig.(\ref{figb}d) we show a ``phase diagram" of different CHS defined by Eq.(\ref{ham}) with varying values of $\alpha$ and $\nu$, and for simplicity we focus on $\nu\le 1$. The blue region are the CHS within the LLL: $L_1$ is the entire LLL, $L_2$ is the null space of $V^{\text{2bdy}}_0$ (Laughlin $1/2$ ground states and quasiholes), and $L_3$ is the null space of $V^{\text{2bdy}}_0+V^{\text{2bdy}}_2$ (Laughlin $1/4$ ground states and quasiholes). As the filling factor increases, more LLs are involved with $\alpha\ge 2$: both $S_1$ and $S_2$ are CHS involving two lowest LLs, and each is a \emph{subspace} of two LLs but isomorphic to the Hilbert space of two LLs with a rescaling of the magnetic field. Similarly, $T_1$ is a CHS involving three lowest LLs, but it is non-Abelian\cite{seidel}.

It is interesting to note that within the red and green region the dominant dynamic energy scale is the kinetic energy. In Fig.(\ref{figa}a) the energy spectra at the filling factor of $\nu=1/2$ on the spherical geometry, with the magnetic monopole at the sphere center $2Q=2N_b-2$ ($N_b$ is the Boson number), can also be understood as from the two following Hamiltonians:
\begin{eqnarray}\label{h12}
&&\mathcal H_1=\hat V^{\text{2bdy}}_0,\quad\mathcal H_2=\sum_i\frac{1}{2m}g^{ab}\hat\pi_a\hat\pi_b\label{h2}
\end{eqnarray}
where $\mathcal H_1$ gives $V^{\text{2bdy}}_0$ PP interaction within $L_1$ (the LLL), and $\mathcal H_2$ is within $S_1$ (the null space of the TK Hamiltonian for the lowest two LLs). Apparently $\mathcal H_1$ is a purely interacting Hamiltonian between Bosons, while $\mathcal H_2$ only involves kinetic energy of the Bosons. Naively it would be surprising to find these two Hamiltonians having very similar spectrum, reflecting a \emph{duality} between the strongly interacting and weakly or non-interacting systems. Now we understand this duality is natural: $\mathcal H_1$ is the special case of Eq.(\ref{ham}) with $\alpha=2-\epsilon$ with $\lim\epsilon\to 0$ in the limit of the large $\omega_c$, while $\mathcal H_2$ is the special case of $\alpha=2$; the two Hamiltonians are very close in the parameter space of $\alpha$.

The dispersion of the MR modes at $\alpha=2, \nu=1/2$ indicates non-trivial interaction between the Laughlin quasihole and quasielectron emerging from the kinetic energy of the Bosons. This is a generic feature of the quasiparticles in the multiple LL CHS as the null space of the interaction between the elementary particles: these quasiparticles are non-interacting without the kinetic energy of the elementary particles; they gain self-energy and start to interact only when the kinetic energy is included. Such energies come from the mixing of the higher LLs for the many-body wavefunctions. The amount of mixing depends on the separation between quasiparticles, which is worth further detailed study as an unusual microscopic mechanism for quasiparticle interaction.

\textit{Experimental platform for non-Abelian phases--} It is still very challenging to realise non-Abelian topological phases in FQH systems. There are tentative results for the Moore-Read phase at $\nu=5/2$\cite{willet}, but the physical interaction (the second LL Coulomb interaction) is very different from the model Hamiltonian consisting of a pure 3-body interaction\cite{read, footnote3}. Ideally, we would like to experimentally realise model Hamiltonians for the non-Abelian phases, for example the two-body TK Hamiltonian within the lowest three or more Landau levels\cite{seidel,ortiz}. The scale-free interaction in Eq.(\ref{ham}) offers exactly that path, but it seems difficult to increase $\alpha$ in a realistic experiment in a 3-dimensional universe. 

Nevertheless, the key for realising CHS involving multiple LLs with strong LL mixing is the competition between the kinetic energy and the energy scale given by \emph{individual two-body pseudopotentials}. In this way, LL mixing explicitly depends on the filling factor. It is important to note that a \emph{strong magnetic field or large $\omega_c$} is needed, because we only hope to mix, or project into, a small number of lower LLs. 

While Fermionic systems are possible, we now focus on Bosonic systems because for Coulomb based interaction, $V^{\text{2bdy}}_0$ is more dominant against other even PPs, while $V^{\text{2bdy}}_1$ is less dominant against other odd PPs in comparison. This is desirable, because a good approximation to the scale-free interaction at $\alpha=2$ is for $||\hat V^{\text{2bdy}}_0||\gg n\hbar\omega_c\gg||\hat V^{\text{2bdy}}_{i>0}||$\cite{footnote2}. We can enhance the interaction effect for particles with greater charge or mass. The main challenge is thus to tune the Coulomb interaction experimentally to achieve the regime where $||\hat V^{\text{2bdy}}_0||\gg||\hat V^{\text{2bdy}}_{i>0}||$.
\begin{figure}
\begin{center}
\includegraphics[width=\linewidth]{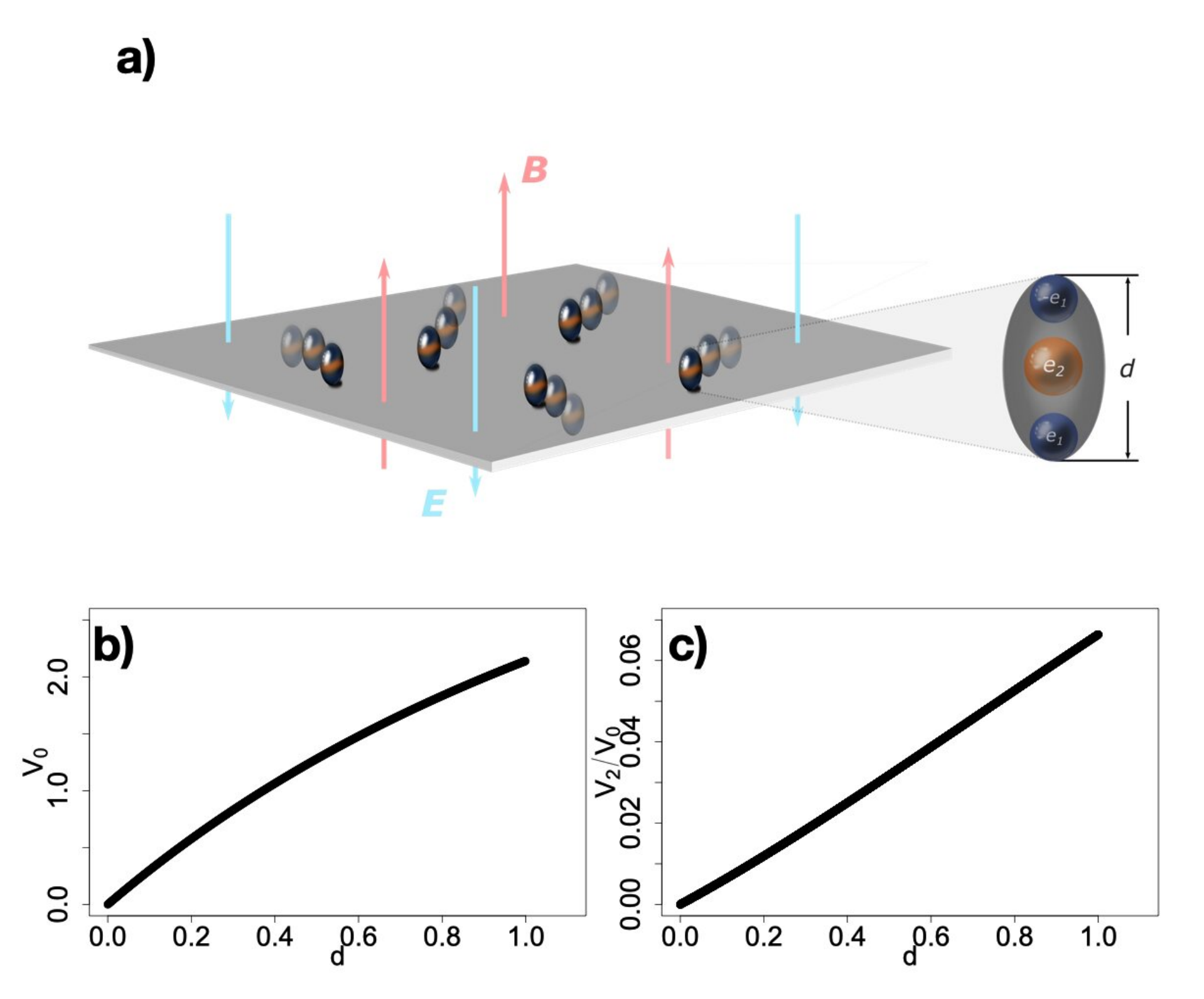}
\caption{a). A schematic illustration of charged Bosons with finite dipole moments confined to a 2D manifold. The electric field is needed to align all the dipoles and it can be either pointing up or pointing down; b). The dependence of $V_0^{\text{2bdy}}$ PP weight on $d$ parameterising the size of the dipole; c). The dependence of the ratio of the PP weights of $V_0^{\text{2bdy}}$ and $V_2^{\text{2bdy}}$ on $d$.}
\label{figc}
\end{center}
\end{figure}

Here we propose a ``simple" physical system in which the required window for different energy scales can be readily tuned and achieved. It offers a novel pathway for realising approximate model Hamiltonians for (non-Abelian) FQH phases. We hypothesise a collection of \emph{charged Bosons with a non-zero dipole moment} confined to a two-dimensional manifold subject to a strong perpendicular magnetic field (see Fig.(\ref{figc}a)), and an out-of-plane electric field to align their dipole moment. They can be considered as Bosonic analogous of ``trions" that have been realised in certain semiconductor systems\cite{xiaodong,shanjie,dery}. Let the total dipole and charge of each Boson be $e_1d$ and $e_2$ respectively. The magnetic length and the cyclotron energy are given by $\ell_B=\sqrt{\hbar/e_2B}$ and $\epsilon_c=\frac{\hbar e_2B}{m}$ (where $m$ is the Boson mass). The Boson-Boson interaction is:
\begin{eqnarray}  
&&V\left(\bm r\right)=\frac{1}{4\pi\epsilon}\left(\epsilon_1\left(\bm r\right)+\epsilon_2\left(\bm r\right)\right)\\
&&\epsilon_1=2e_1^2\left(\frac{1}{\bm r}-\frac{1}{\sqrt{\bm r^2+d^2}}\right), \epsilon_2=\frac{e_2^2}{\bm r}
\end{eqnarray}
where $\bm r=|\vec r|$. Note that for $\bm r\gg d$ we have $\epsilon_1\sim \bm r^{-3}$, but $\epsilon_1$ vanishes as $d\to 0$, thus we do not have scale-free interaction at short distances (see Fig.(\ref{figc}b)). 

Nevertheless the magnetic length $\ell_B$ sets a ``minimum length" and we can decompose $V\left(\bm r\right)$ into PPs within the LLL with weights $c_i$. It is not surprising that we find $\lim_{d\to 0}c_{i>0}/c_0=0$, so that $V^{\text{2bdy}}_0$ dominates (see Fig.(\ref{figc}c)). For large magnetic field we have $\epsilon_c\gg\epsilon_1$, thus here we ignore $\epsilon_1$. Assuming we can tune $e_1, e_2$ and $m$ of the charged Bosons, we can then get arbitrarily close to the regime with $e_1\gg e_2/m$. It is thus potentially feasible to engineer the 2D system to realise many interesting (non-Abelian) FQH phases (e.g. the bosonic Jain 221 state with 3 LLs and the Fibonacci state with 4 LLs).

One focus of the future studies is the anyon interactions within multiple LLs emerging from the kinetic energy. When the energy window proposed above is achieved, we can realise robust ground states that are theoretically proposed with prominent Hall plateaus at those Abelian and non-Abelian phases. However, the quasihole excitations may not be degenerate in the thermodynamic limit due to the cyclotron energy. Numerical analysis of multi-LL systems, however, is strongly constrained by the large size of the Hilbert space. Thus better numerical and theoretical tools are needed to understand the interesting dynamics (both gapped and gapless) for these topological phases.


\begin{acknowledgments}
{\sl Acknowledgements.} I would like to acknowledge useful discussions with A.C. Balram and G.J. Sreejith, and help from Yuzhu Wang for the figures. This work is supported by the NTU grant for Nanyang Assistant Professorship and the National Research Foundation, Singapore under the NRF fellowship award (NRF-NRFF12-2020-005), and Singapore Ministry of Education (MOE) Academic Research Fund Tier 3 grant (MOE-MOET32023-0003) “Quantum Geometric Advantage”.
\end{acknowledgments}

\onecolumngrid
\pagebreak
\widetext
\begin{center}
\textbf{\large Supplementary Online Materials for ``Emergent dynamics of scale-free interactions in fractional quantum Hall fluids"}
\end{center}
\begin{center}
Bo Yang\\
\textit{Division of Physics and Applied Physics, Nanyang Technological University, Singapore 637371.}
\end{center}
\setcounter{equation}{0}
\setcounter{figure}{0}
\setcounter{table}{0}
\setcounter{page}{1}
\makeatletter
\renewcommand{\theequation}{S\arabic{equation}}
\renewcommand{\thefigure}{S\arabic{figure}}
\renewcommand{\bibnumfmt}[1]{[S#1]}
\renewcommand{\citenumfont}[1]{S#1}
In this supplementary material, we give more details on the dynamical properties of the scale free interactions, as well as the decomposition of such interactions in the two-body pseudopotential basis.

\section{S1. Energy spectrum of the scale free interactions}

\begin{figure}[h!]
\begin{center}
\includegraphics[width=\linewidth]{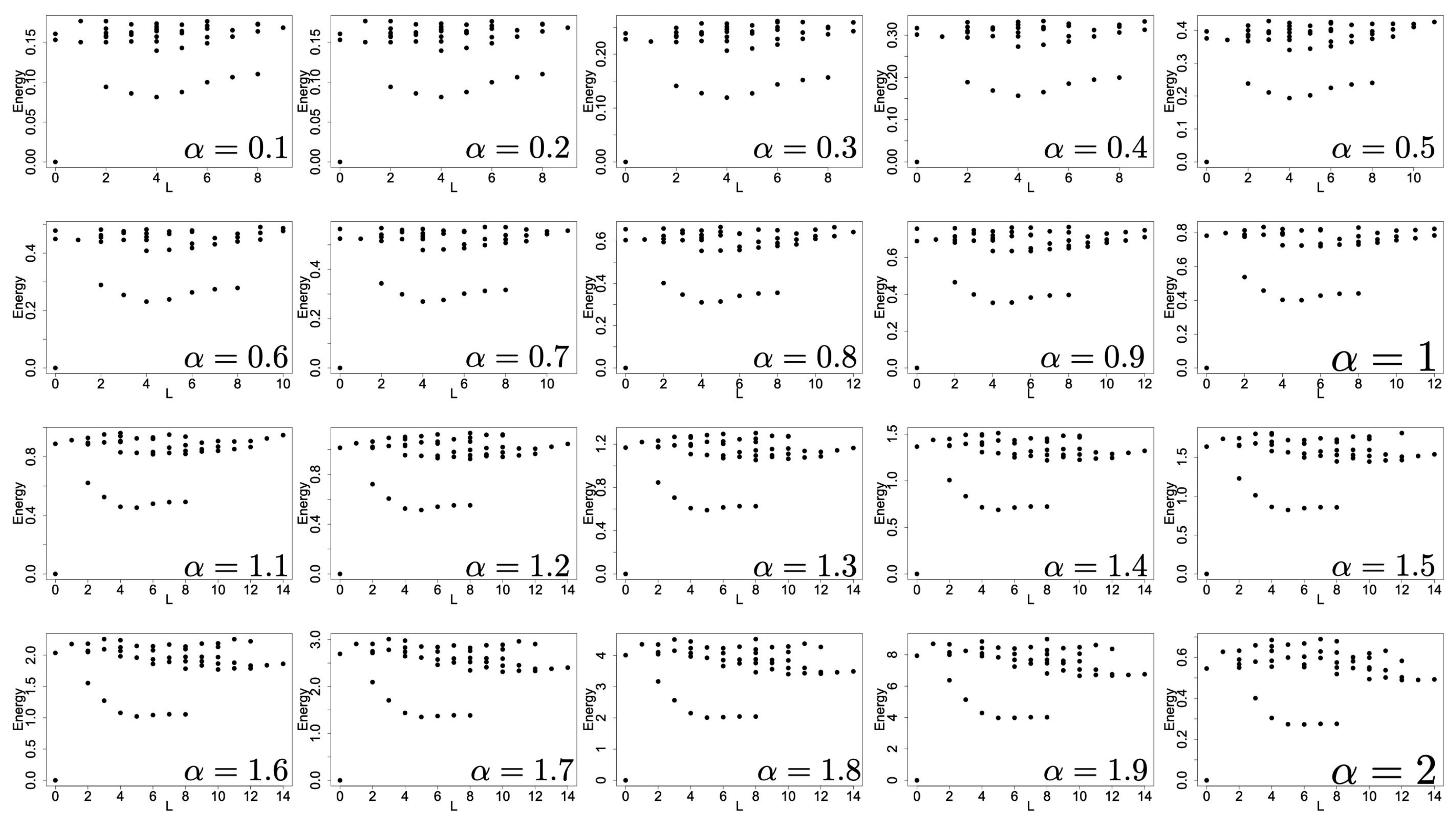}
\caption{The energy spectra for the system of $8$ bosons in $15$ orbitals, for various different values of $\alpha$ as indicated in the plots.}
\label{s1}
\end{center}
\end{figure}

We present here more details on the dynamical properties of the family of the scale free interactions. In the main text we have shown the energy spectrum at filling factor $\nu=1/2$ for only a few values of $\alpha$ with the two-body interaction $V_\alpha(r)=1/r^\alpha$. In Fig.(\ref{s1}), the energy spectra for all values of $\alpha$ between $0$ and $2$ are presented (with the incremental step of $\delta\alpha=0.1$). We can see the spectra are all qualitatively the same, even though in the language of pseudopotentials, the two-body interaction changes drastically. The overlap between the ground state and the Laughlin model wavefunction can also be shown in Fig.(\ref{s2}), indicating excellent overlap for all values of $\alpha$. Naively speaking this is quite unexpected, as for example at $\alpha=0.1$ the pseudopotential components are $c_0=0.97, c_2=0.9, c_3=0.88. c_4=0.87,\cdots$. The very large $c_{i>0}$ as compared to $c_0$ normally suggests that the ground state should be very different from the Laughlin model state, and we would not even expect it to appear in the $L=0$ sector. As one can see in Fig.(\ref{s2}b), at $\alpha=0.1$ the overlap with the model state is extremely high even for the very large system of 12 particles.

\begin{figure}[h!]
\begin{center}
\includegraphics[width=\linewidth]{s2.pdf}
\caption{The overlap of the $V=1/r^\alpha$ ground state with the model Laughlin $\nu=1/2$ state a) for different values of $\alpha$ with $8$ particles; b). at $\alpha=0.1$ for $8, 10$ and $12$ particles.}
\label{s2}
\end{center}
\end{figure}

\section{S2. Residual pseudopotentials for the $\alpha=2$ scale free interaction}
As explained in the main text, for scale free interactions with $\alpha\ge 2$, the leading pseudopotentials will diverge, while the remaining pseudopotentials are finite and small. The diverging pseudopotentials correspond exactly to the non-zero pseudopotentials from the corresponding TK Hamiltonian. In this section, using $\alpha=2$ as an example, we present a few significant pseudopotentials for two-particle states up to the third Landau level. Let the two-particle states be denoted as $|m,n_1,n_2\rangle$, where $m$ is the relative angular momentum between the two particles that are in the $n_1^{\text{th}}$ and $n_2^{\text{th}}$ Landau level (LL) respectively. Starting from $|\psi_0\rangle\vcentcolon=|\psi_{m,0}\rangle = |m,00\rangle$, which denotes a pair of particles both in the LLL with relative angular momentum $m$, the complete basis up to the second LL are given as follows:
\begin{eqnarray}
&&|\psi_1\rangle\vcentcolon=|\psi_{m,1}\rangle\sim a_{1,2}^\dagger|\psi_{m+1,0}\rangle\\
&&|\psi_2\rangle\vcentcolon=|\psi_{m,2}\rangle\sim A_{1,2}^\dagger|\psi_{m,0}\rangle\\
&&|\psi_3\rangle\vcentcolon=|\psi_{m,3}\rangle\sim \left(a_{1,2}^\dagger\right)^2|\psi_{m+2,0}\rangle\\
&&|\psi_4\rangle\vcentcolon=|\psi_{m,4}\rangle\sim \left(A_{1,2}^\dagger\right)^2|\psi_{m,0}\rangle\\
&&|\psi_5\rangle\vcentcolon=|\psi_{m,5}\rangle\sim A_{1,2}^\dagger a_{1,2}^\dagger|\psi_{m+1,0}\rangle\\
&&|\psi_6\rangle\vcentcolon=|\psi_{m,6}\rangle\sim a_1^\dagger a_2^\dagger(a_1^\dagger-a_2^\dagger)|\psi_{m+3,0}\rangle\sim\left(\left(A_{1,2}^\dagger\right)^2a_{1,2}^\dagger-\left(a_{1,2}\right)^3\right)|\psi_{m+3,0}\rangle\\
&&|\psi_7\rangle\vcentcolon=|\psi_{m,7}\rangle\sim a_1^\dagger a_2^\dagger(a_1^\dagger+a_2^\dagger)|\psi_{m+2,0}\rangle\sim\left(\left(A_{1,2}^\dagger\right)^3-\left(a_{1,2}\right)^2A_{1,2}^\dagger\right)|\psi_{m+2,0}\rangle\\
&&|\psi_8\rangle\vcentcolon=|\psi_{m,8}\rangle\sim \left(a_1^\dagger\right)^2\left(a_2^\dagger\right)^2|\psi_{m+4,0}\rangle\sim\left(\left(A_{1,2}^\dagger\right)^4+\left(a_{1,2}^\dagger\right)^4-2\left(A_{1,2}^\dagger\right)^2\left(a_{1,2}^\dagger\right)^2\right)|\psi_{m+4,0}\rangle
\end{eqnarray}
where $a_{1,2}^\dagger\sim a_1^\dagger-a_2^\dagger, A_{1,2}^\dagger\sim a_1^\dagger+a_2^\dagger$ are the raising operators to higher LLs with $1,2$ as the particle index; $|\psi_{m,i}\rangle$ with $i=1,2,3,4,5$ has relative angular momentum $m$, $|\psi_{m,i}\rangle$ with $i=6,7$ is a linear combination of states with relative angular momentum $m$ and $m+2$, while $|\psi_{m,8}\rangle$ is a linear combination with relative angular momentum $m,m+2,m+4$. The coefficients of the pseudopotentials can thus be readily computed as follows
\begin{eqnarray}
c_{\alpha,m}^{ij}=\langle \psi_i|\hat V_\alpha|\psi_j\rangle
\end{eqnarray}
via real space integration, where in the real basis $\hat V_\alpha\sim 1/r^\alpha$. For example the coefficients of the familiar pseudopotentials (i.e. model Hamiltonians for the Laughlin states) in the LLL are given by $c_{\alpha,m}^{00}$. The coefficients are only non-vanishing if $|\psi_i\rangle$ and $|\psi_j\rangle$ have the same relative angular momentum and contain the same power of $A_{1,2}^\dagger$. We list here the analytic expressions for these nonzero coefficients:
\begin{eqnarray}
&&c_{\alpha,m}^{00}=\frac{2^{-\alpha}\Gamma(1+m-\alpha/2)}{m!}\label{s}\\
&&c_{\alpha,m}^{01}=-\frac{2^{-\alpha}\alpha\Gamma(1+m-\alpha/2)}{\sqrt{m!(m+1)!}}\\
&&c_{\alpha,m}^{03}=\frac{2^{-3/2-  \alpha} \alpha (1 + \alpha/2) \Gamma(1 + m - \alpha/2)}{\sqrt{m!(m+2)!}}\\
&&c_{\alpha,m}^{06}=\frac{2^{-5/2 -  \alpha} \alpha (1 + \alpha/2)(2+\alpha/2)  \Gamma(  1 + m - \alpha/2)}{\sqrt{m!(m+3)!}}\\
&&c_{\alpha,m}^{08}=\frac{2^{-4- \alpha} \alpha (1 + \alpha/2) (2 + \alpha/2)(3+\alpha/2) \Gamma(1 + m - \alpha/2)}{\sqrt{m!(m+4)!}}\\
&&c_{\alpha,m}^{11}=\frac{2^{-  \alpha}(m+ (-1 + \alpha/2) \alpha/2) \Gamma(m - \alpha/2)}{m!}\\
&&c_{\alpha,m}^{13}=-\frac{2^{-3/2- \alpha} \alpha (2 m + (-1 + \alpha/2) \alpha/2) \Gamma(m - \alpha/2)}{\sqrt{m!(m+1)!}}\\
&&c_{\alpha,m}^{16}=-\frac{2^{-5/2 -\alpha} \alpha (1 + \alpha/2) (3 m + (-1 + \alpha/2) \alpha/2)  \Gamma(m - \alpha/2)}{\sqrt{m!(m+2)!}}\\
&&c_{\alpha,m}^{18}=-\frac{2^{-4 - \alpha} \alpha (1 + \alpha/2) (2 + \alpha/2) (4 m + (-1 + \alpha/2) \alpha/2) \Gamma(m - \alpha/2)}{\sqrt{m!(m+3)!}}\\
&&c_{\alpha,m}^{22}=\frac{2^{-  \alpha}\Gamma(m+1 - \alpha/2)}{m!}\\
&&c_{\alpha,m}^{25}=-\frac{2^{-1/2 -  \alpha} \alpha\Gamma( m+1 - \alpha/2)}{\sqrt{(m+1)!m!}}\\
&&c_{\alpha,m}^{27}=-\frac{2^{-5/2-\alpha}\alpha (1 + \alpha/2)  \Gamma(1 + m - \alpha/2)}{\sqrt{m!(m+2)!}}\\
&&c_{\alpha,m}^{33}=\frac{2^{-1-\alpha}(2 m^2 + (-2 + \alpha/2) (-1 + \alpha/2) \alpha/2 (1 + \alpha/2) + m (-2 + 4 (-1 + \alpha/2) \alpha/2)) \Gamma(-1 + m - \alpha/2)}{m!}\\
&&c_{\alpha,m}^{36}=\frac{2^{-3 - \alpha} \alpha (6 m^2 + (-2 + \alpha/2) (-1 + \alpha/2) \alpha/2 (1 + \alpha/2) + 6 m (-1 + (-1 + \alpha/2) \alpha/2))  \Gamma(-1 + m - \alpha/2)}{\sqrt{(m+1)!m!}}\\
&&c_{\alpha,m}^{38}=\frac{2^{-9/2 - \alpha} \alpha (1 + \alpha/2) (6 m + (-1 + \alpha/2) \alpha/2) (2 m + (-2 + \alpha/2) (1 + \alpha/2))  \Gamma(-1 + m - \alpha/2)}{\sqrt{m!(m+2)!}}\\
&&c_{\alpha,m}^{44}=\frac{2^{- \alpha}\Gamma( 1+m - \alpha/2)}{m!}\\
&&c_{\alpha,m}^{46}=-\frac{2^{-2 - \alpha} \alpha\Gamma(1 + m - \alpha/2)}{\sqrt{m!(m+1)!}}\\
&&c_{\alpha,m}^{55}=\frac{2^{ -  \alpha} (m+ (-1 + \alpha/2) \alpha/2) \Gamma(m - \alpha/2)}{m!}\\
&&c_{\alpha,m}^{57}=\frac{2^{-5/2- \alpha}\alpha (2 m + (-1 + \alpha/2) \alpha/2)  \Gamma(m - \alpha/2)}{\sqrt{m!(m+1)!}}\\
&&c_{\alpha,m}^{66}=\frac{2^{-3 -  \alpha}}{m!}\Bigg(8 m^3 + (-1 + \alpha/2) \alpha/2 (1 + \alpha/2) (2 + \alpha/2) (8 + (-5 + \alpha/2) \alpha/2) + \nonumber\\
&&\qquad\qquad\qquad4 m^2 (-6 + \alpha/2 (-6 + 5 \alpha/2)) + \nonumber\\
&&\qquad\qquad\qquad m (16 + \alpha/2 (48 + \alpha/2 (-27 + \alpha/2 (-22 + 9 \alpha/2))))) \Gamma(-2 + m -\alpha/2)\Bigg)\\
&&c_{\alpha,m}^{68}=\frac{2^{-11/2-\alpha}}{\sqrt{m!(m+1)!}}\Bigg(\alpha (32 m^3 +8 m^2 (1 + \alpha/2) (-12 + 5 \alpha/2) +\nonumber\\
&&\qquad\qquad\qquad\qquad\qquad (-1 + \alpha/2) \alpha/2 (1 + \alpha/2) (2 + \alpha/2) (10 + (-5 + \alpha/2) \alpha/2) + \nonumber\\
&&\qquad\qquad\qquad\qquad\qquad 4 m (-4 + (-1 + \alpha/2) \alpha/2) (-4 + \alpha/2 (-5 + 3 \alpha/2)))  \Gamma(-2 + m - \alpha/2)\Bigg)
\end{eqnarray}
\begin{eqnarray}
&&c_{\alpha,m}^{77}=\frac{2^{-3- \alpha} (8 m^2 + 4 m (-2 + (-4 + \alpha/2) \alpha/2) + \alpha/2 (1 + \alpha/2) (8 + (-3 + \alpha/2) \alpha/2)) \Gamma(-1 + m - \alpha/2)}{m!}\quad\quad\\
&&c_{\alpha,m}^{88}=\frac{2^{-6-\alpha}}{m!} \Bigg((64 m^4 + 128 m^3 (-3 - 2 \alpha/2 + \alpha^2/4) +16 m^2 (44 + 36 \alpha - 19 \alpha^2/4 - 7 \alpha^3/4 + 5 \alpha^4/16)\nonumber\\
&&\qquad\qquad\qquad\qquad+ 16 m (-24 - 44 \alpha - 5 \alpha^2/4 + 4 \alpha^3 -  \alpha^4/2 -  \alpha^5/8 + \alpha^6/64) +\nonumber\\
&&\qquad\qquad\qquad \alpha/2(384 + 130 \alpha - 43 \alpha^2 + 33 \alpha^3/8 + 5 \alpha^4 - 3 \alpha^5/16 - \alpha^6/16 + \alpha^7/128)) \Gamma(-3 + m - \alpha/2)\Bigg)\label{ss}
\end{eqnarray}

The explicit values of these coefficients are shown in Table.\ref{t1}, for the Coulomb interaction $\alpha=1$ and the specific case of $\alpha=2$, where we set the magnetic length $\ell_B=1$. For the non-diverging pseudopotentials with $\alpha=2$, they are sometimes comparable, and generally significantly smaller than those from the Coulomb interactions. One can easily check that for $\alpha>2$, the non-diverging pseudopotentials are even smaller as compared to the Coulomb interaction.

It is thus important to note that while we refer to the ``null spaces" of scale free interactions in the limit of infinite cyclotron energy in the main text, these are spaces where the states still have finite energies due to the finite pseudopotentials (e.g. listed in Table.~\ref{t1}), which are taken as ``zero energies" in comparison to the infinite energy scales from the diverging pseudopotentials and the cyclotron gap. The eigenstates of the scale free interactions are exact model ground states and quasihole states, but the anyons in the null space will start interacting due to the finite pseudopotentials. Such interaction is not important for a dilute anyon gas, or when the separation of the two anyons is large (a necessary condition for anyon braiding even for non-interacting anyons).

\begin{table}[ht]
\begin{tabular}{|c|c|c|c|}
 \hline
  &Leading relevant $m$ &$\alpha=1$& $\alpha=2$ \\ 
 \hline
\rule{0pt}{4ex}$c_{\alpha,m}^{00}$&$m=0,2,4$  & 0.886, 0.332, 0.242 & $\infty$, 0.125, 0.063\\ [8pt]
 \hline
\rule{0pt}{4ex}$c_{\alpha,m}^{01}$ &$m=0,2,4$ & -0.443, -0.096, -0.054 & $-\infty$, -0.072, -0.028\\[8pt]
\hline
\rule{0pt}{4ex}$c_{\alpha,m}^{03}$ &$m=0,2,4$ & 0.332, 0.051, 0.023 & $\infty$, 0.051, 0.016\\  [8pt]
 \hline
\rule{0pt}{4ex}$c_{\alpha,m}^{06}$  &$m=0,2,4$& 0.24, 0.028, 0.011 &$\frac{\sqrt 3}{2}\infty$, 0.034, 0.009\\  [8pt]
 \hline
\rule{0pt}{4ex}$c_{\alpha,m}^{08}$  &$m=0,2,4$& 0.148, 0.014, 0.005 & $\frac{\sqrt 6}{4}\infty$, 0.02, 0.005\\  [8pt]
 \hline
\rule{0pt}{4ex}$c_{\alpha,m}^{11}$ &$m=0,2,4$ & 0.665, 0.305, 0.23 & $\infty$, 0.125, 0.063\\  [8pt]
 \hline
\rule{0pt}{4ex}$c_{\alpha,m}^{13}$& $m=0,2,4$ & -0.388, -0.112, -0.068 & $-\infty$, -0.088, -0.036\\  [8pt]
 \hline
\rule{0pt}{4ex}$c_{\alpha,m}^{16}$ &$m=0,2,4$ & -0.264, -0.057, -0.029 &$-\frac{\sqrt 3}{2}\infty$, -0.059, -0.02 \\  [8pt]
 \hline
\rule{0pt}{4ex}$c_{\alpha,m}^{18}$  &$m=0,2,4$& -0.159, -0.028, -0.012 & $-\frac{\sqrt 6}{4}\infty$, -0.034, -0.01\\[8pt]  
 \hline
 \rule{0pt}{4ex}$c_{\alpha,m}^{22}$ &$m=0,2,4$ & 0.886, 0.332, 0.242 & $\infty$, 0.125, 0.063\\  [8pt]
  \hline
\rule{0pt}{4ex}$c_{\alpha,m}^{25}$ &$m=0,2,4$ & -0.443, -0.096, -0.054 & $-\infty$, -0.072, -0.028\\  [8pt]
  \hline
\rule{0pt}{4ex}$c_{\alpha,m}^{27}$ & $m=0,2,4$& -0.166, -0.025, -0.012 & $-\frac{1}{2}\infty$, -0.026, -0.008\\  [8pt]
 \hline
\rule{0pt}{4ex}$c_{\alpha,m}^{33}$ &$m=0,2,4$ & 0.332, 0.568, 0.285 &0.125, $\infty$, 0.125\\  [8pt]
 \hline
\rule{0pt}{4ex} $c_{\alpha,m}^{36}$ &$m=0,2,4$ & 0.083, 0.306, 0.103 &0.063, $\frac{\sqrt 3}{2}\infty$, 0.084\\  [8pt]
 \hline
\rule{0pt}{4ex} $c_{\alpha,m}^{38}$ & $m=0,2,4$& 0.031, 0.174, 0.046 & 0.031,$\frac{\sqrt 6}{4}\infty$, 0.048\\  [8pt]
 \hline
\rule{0pt}{4ex}  $c_{\alpha,m}^{44}$  &$m=0,2,4$& 0.886, 0.332, 0.242 & $\infty$, 0.125, 0.063\\  [8pt]
 \hline
\rule{0pt}{4ex}  $c_{\alpha,m}^{46}$ &$m=0,2,4$ & -0.222, -0.048, -0.027 & $-\frac{1}{2}\infty$, -0.036, -0.014\\[8pt]  
 \hline
 \rule{0pt}{4ex} $c_{\alpha,m}^{55}$&$m=0,2,4$  & 0.665, 0.305, 0.23 & $\infty$, 0.125, 0.063\\  [8pt]
 \hline
\rule{0pt}{4ex}  $c_{\alpha,m}^{57}$ &$m=0,2,4$ & 0.194, 0.056, 0.034 & $\frac{1}{2}\infty$, 0.044, 0.018\\ [8pt] 
 \hline
\rule{0pt}{4ex}  $c_{\alpha,m}^{66}$ &$m=0,2,4$ & 0.395, 0.458, 0.26 & $\frac{1}{4}\infty$, $\frac{3}{4}\infty$, 0.109\\  [8pt]
 \hline
\rule{0pt}{4ex}  $c_{\alpha,m}^{68}$ &$m=0,2,4$ & 0.157, 0.202, 0.082 & $\frac{1}{4}\infty$, $\frac{3\sqrt 2}{8}\infty$, 0.063\\  [8pt]
 \hline
\rule{0pt}{4ex}   $c_{\alpha,m}^{77}$ &$m=0,2,4$ & 0.748, 0.391, 0.253 & $\frac{3}{4}\infty$, $\frac{1}{4}\infty$, 0.078\\  [8pt]
 \hline
\rule{0pt}{4ex}   $c_{\alpha,m}^{88}$& $m=0,2,4,6$ & 0.506, 0.373, 0.337, 0.226 & $\frac{3}{8}\infty$, $\frac{1}{4}\infty$, $\frac{3}{8}\infty$, 0.078\\ [8pt] 
 \hline
\end{tabular}
\caption{Some values of the pseudopotential components (relevant to the bosons) for the scale free interactions at $\alpha=1$ and $\alpha=2$, from the analytic expressions from Eq.(\ref{s}) to Eq.(\ref{ss}).}
\label{t1}
\end{table}


\begin{thebibliography}{99}
\bibitem{footnote} Note that in the second LL (1LL), $\alpha=2$ interaction has infinite $V_0$ and $V_2$ pseudopotentials. Together with interaction of particles between the LLL and 1LL, we can only fill the 1LL up to the filling fraction of $1/6$ (thus total filling of $1/2+1/6=2/3$ before having to put particles into the third LL.
\bibitem{footnote1} The manual selection of LLs could be physical for multilayer graphene where multiple LLs are degenerate\cite{yinghai,mccann,macdonald}. Such interesting systems have their own caveats and we will not focus on those here.
\bibitem{fn} For $\alpha=2,4$, the bosonic and fermionic versions of Jain state (with 2LLs), the non-Abelian Jain 221 state (with 3LLs) and Jain 222 state (with 4 LLs) are realised. Other verified and conjectured states involving more LLs can be found in Table 5 of Ref.\cite{ortiz}. For larger values of $\alpha$, the physically realised (non-Abelian) topological phases are related to those conjectured in Ref.\cite{wenxg}.
\bibitem{footnote5} This is also true for other combinations of $2Q$ and $N_b$, with the presence of additional quasiholes and quasielectrons.
\bibitem{footnote3} With Coulomb interaction it is not clear if the $1/4$ charge quasiparticles are energetically favourable, or the degeneracy required for non-Abelian statistics is robust in experiments, even though we observe an incompressible ground state leading to a Hall plateau at $\nu=5/2$.
\bibitem{footnote2} Here $||\cdots||$ indicates the energy scale of the PP interaction component. For a generic two-body interaction Hamiltonian we have $\hat V_q=\sum_i\hat V_i^{\text{2bdy}}$, which is how $\hat V_i^{\text{2bdy}}$ is defined; $n$ is the number of the lowest LLs that intends to be mixed
\end{thebibliography}
\end{document}